\documentclass[aip,cha,preprint,numerical]{revtex4-1}

\usepackage[T2A]{fontenc}
\usepackage[utf8]{inputenc}
\usepackage{amsmath}
\usepackage{amssymb}
\usepackage{graphicx}
\usepackage{xcolor}
\usepackage{natbib}
\usepackage{bm}

\begin{document}

\title{Appearance of chaos and hyperchaos  in evolving pendulum network}

	\author{Vyacheslav O. Munyaev}
	\affiliation{Department of Control Theory, Scientific and Educational Mathematical Center ``Mathematics of Future Technologies'', Nizhny Novgorod State University, Gagarin Av. 23, Nizhny Novgorod, 603950 Russia}
	
	\author{Dmitry S. Khorkin}
	\affiliation{Department of Control Theory, Scientific and Educational Mathematical Center ``Mathematics of Future Technologies'', Nizhny Novgorod State University, Gagarin Av. 23, Nizhny Novgorod, 603950 Russia}
	
	\author{Maxim I. Bolotov}
	\affiliation{Department of Control Theory, Scientific and Educational Mathematical Center ``Mathematics of Future Technologies'', Nizhny Novgorod State University, Gagarin Av. 23, Nizhny Novgorod, 603950 Russia}
	
	\author{Lev A. Smirnov}
	\affiliation{Department of Control Theory, Scientific and Educational Mathematical Center ``Mathematics of Future Technologies'', Nizhny Novgorod State University, Gagarin Av. 23, Nizhny Novgorod, 603950 Russia}
	\affiliation{Institute of Applied Physics, Russian Academy of Sciences, Ul’yanova Str. 46, Nizhny Novgorod, 603950 Russia}
	
	\author{Grigory V. Osipov}
	\affiliation{Department of Control Theory, Scientific and Educational Mathematical Center ``Mathematics of Future Technologies'', Nizhny Novgorod State University, Gagarin Av. 23, Nizhny Novgorod, 603950 Russia}
\begin{abstract}
The study of deterministic chaos continues to be one of the important problems in the field of nonlinear dynamics. Interest in the study of chaos exists both in low-dimensional dynamical systems and in large ensembles of coupled oscillators. In this paper, we study the emergence of spatio-temporal chaos in chains of locally coupled identical pendulums with constant torque. The study of the scenarios of the emergence (disappearance) and properties of chaos is done as a result of changes in: (i) the individual properties of elements due to the influence of dissipation in this problem, and (ii) the properties of the entire ensemble under consideration, determined by the number of interacting elements and the strength of the connection between them. 
It is shown that an increase of dissipation in an ensemble with a fixed coupling force and elements number can lead to the appearance of chaos as a result of a cascade of period doubling bifurcations of periodic rotational motions or as a result of invariant tori destruction bifurcation. Chaos and hyperchaos can occur in an ensemble by adding or excluding one or more elements. Moreover, chaos arises hard, since in this case the control parameter is discrete. The influence of the coupling strength on the occurrence of chaos is specific. The appearance of chaos occurs with small and intermediate coupling and is caused by the overlap of the various out-of-phase rotational modes regions existence. The boundaries of these areas are determined analytically and confirmed in a numerical experiment. Chaotic regimes in the chain do not exist if the coupling strength is strong enough.
\end{abstract}
\date{\today}
\date{\today}
\maketitle
	\begin{quotation}
        Networks of interacting nonlinear oscillators are encountered in various natural and technical situations. They  govern the behavior of coupled neurons and cardiac cells, many physical devices such as arrays  of Josephson junctions and lasers, many engineering applications such as phase locked loops and electric power machines. Because of many applications, the study of collective dynamics, especially synchronization spatio-temporal chaos  are some  of central subjects in nonlinear dynamics for the past three decades. Many significant and important results has been obtained. In our paper we present a theoretical and computational study of complex dynamics in the chains of coupled pendulums. The influence of individual  (dissipation of pendulums) and collective (coupling strength and number of elements) properties of population is analyzed and discussed.
	\end{quotation}
\maketitle
\section{Introduction}\label{sec:Introduction}
The study of spatio-temporal dynamics in ensembles of nonlinear oscillators of various nature is one of the most popular and interesting directions in modern nonlinear dynamics. The behavior of elements of such ensembles can be roughly divided into three types: (i) fully organized (consistent, coherent, synchronous), (ii) completely disorganized (inconsistent, incoherent, asynchronous), and (iii) intermediate between (i) and (ii) (partially consistent, partially coherent, partially synchronous, such as chimeric or cluster states). The most difficult variant of partially synchronous behavior is \textit{spatio-temporal chaos}.
\par
Chaotic oscillations are one of the most common phenomena in nonlinear dynamical systems of dimension three and higher. It can be assumed that the main mechanisms of the appearance of chaotic dynamics are currently well understood.~\cite{Ott-02}. Chaos can be conservative and dissipative. The mathematical image of dissipative chaos is a strange attractor -- nontrivial stable closed invariant set with unstable behavior of trajectories on it. Below we list the main, most important, well-known scenarios of the appearance of strange attractors typical for wide classes of dynamical systems:
\\
(i) bifurcation of the destruction of an invariant torus;
\\
(ii) intermittency;
\\
(iii) infinite sequence of period doubling bifurcations of periodic motions;
\\
(iiii) internal crises of attractors.
\par
All these scenarios of the appearance (and disappearance) of dynamic chaos have been studied in detail for low-dimensional systems~\cite{Afraimovich-83,Pomeau-80,Feigenbaum-80,Grebogi-82,Gonchenko-14}. From the point of view of bifurcation theory, nothing unusual happens when chaos occurs in large distributed systems -- discrete networks and media~\cite{Bohr-98,Holmes-96}. However, it is clear that the behavior of ensembles not only affects the individual dynamics of its components, but also the different characteristics of the interaction of elements: the type and strength of connections, configuration and capacity of the network and others.
\par
In this work, the complication of space-time behavior up to the onset of chaos is investigated in a chain of pendulum-type systems depending on (i) the dissipation of a partial element, (ii) the characteristics of the interaction between the elements -- the strength of connections and (iii) the number of interacting elements.
\par
The work is structured as follows. Section~\ref{sec:Model} describes the studied model of a chain of pendulum elements. In the section~\ref{sec:InPhaseMode}, in-phase rotational motion and the issue of its stability are considered, asymptotic expressions are given for the boundaries of instability regions depending on the coupling strength of the elements in the chain. Section~\ref{sec:UnstabilityRanges} provides a detailed description of possible variants of the in-phase mode unstable regions intersection, depending on the change in the number of elements in the chain. Further, in section~\ref{sec:chaos}, a scenario of the development of chaotic dynamics in a chain with an increase in the dissipation parameter with a fixed number of elements, as well as in the case of an increase in the number of elements with a fixed large dissipation is described. In the Conclusion, the main findings on the presented results of the work are formulated.
\section{Model}\label{sec:Model}
We consider $N$ coupled pendulum-type systems described by a system of ordinary differential equations:
\begin{equation}
\begin{gathered}
\ddot{\varphi}_1+\lambda\dot{\varphi}_1+\sin{\varphi_1}=\gamma+K \sin{\left(\varphi_{2}-\varphi_1\right)},\\
\ddot{\varphi}_n+\lambda\dot{\varphi}_n+\sin{\varphi_n}=\gamma+K \left[\sin{\left(\varphi_{n+1}-\varphi_n\right)} + \sin{\left(\varphi_{n-1}-\varphi_n\right)}\right],\hspace{2mm}n=2,\ldots,N-1,\\
\ddot{\varphi}_{N}+\lambda\dot{\varphi}_N+\sin{\varphi_N}=\gamma+K \sin{\left(\varphi_{N-1}-\varphi_N\right)},
\end{gathered}
\label{eq:ModelGeneral}
\end{equation}
where $\lambda$ -- damping parameter, responsible for dissipative processes in the system, $\gamma$ -- constant torque is the same for all $N$ pendulums, $K$ -- parameter of coupling strength between elements. 
\par
The system~\eqref{eq:ModelGeneral} is used to describe the behavior of interacting pendulums~\cite{Pikovsky-01,Osipov-07,Smirnov-16} (we present our results in the interpretation of pendulums), connected Josephson junctions~\cite{Leeman-86,Kim-95,Denniston-95,Barone-82,Belykh-77_1,Belykh-77_2}. It is used to describe processes in superconductors~\cite{Fishman-88}, molecular biology~\cite{Yakushevich-04,Yakushevich-11},  phase synchronization systems~\cite{Afraimovich-94}. Moreover, such a system can be considered as a generalization of the Kuramoto model taking into account the inertia and intrinsic nonlinearity of the ensemble elements~\cite{Ji-14,Lafuerza-10,Belykh-16,Belykh-20}.
\par
Since the elements of the ensemble are identical, the following spatially homogeneous modes exist for any values of the parameters:
\begin{itemize}
	\item equilibrium state with phase coherent elements $\varphi_1 = \varphi_2 = \ldots = \varphi_N = \overline{\psi} = \text{const}$;
	\item rotational motion of the elements with coherent phase $\varphi_1(t) = \varphi_2(t) = \ldots = \varphi_N(t) \equiv \psi(t)$.
\end{itemize}
These modes satisfy the pendulum equation:
\begin{equation}
	\ddot{\psi} + \lambda\dot{\psi} + \sin \psi = \gamma.
	\label{eq:pend}
\end{equation}
The equation~\eqref{eq:pend} has been carefully examined in~\cite{Andronov-66,Tricomi-33}.
\par
In this study, we are interested in regular and chaotic rotational modes, therefore, we focus on the values of the parameters $\lambda$ and $\gamma$, for which in the system~\eqref{eq:ModelGeneral} there is an in-phase rotation periodic motion -- in-phase regime (IPR), which corresponds to the area on the plane $(\lambda, \gamma)$, bounded below by the Tricomi bifurcation curve~\cite{Tricomi-33,Andronov-66}.
\par
In our previous works, we considered the features of the rotational dynamics of particular variants of the system~\eqref{eq:ModelGeneral}. In~\cite{Smirnov-16,Khorkin-20} two coupled pendulums were investigated for the cases of symmetric and asymmetric coupling, respectively.
In the article~\cite{Bolotov-19} the case of a chain of three locally coupled elements, and in~\cite{Bolotov-20} $N$ globally coupled pendulums were considered.
\section{In-phase regime and its stability}\label{sec:InPhaseMode}
The possibility of realizing rotational regimes of varying degrees of complexity in the system~\eqref{eq:ModelGeneral} is directly related to the issue of stability of the in-phase regime, the development of instability of which leads to the appearance of out-of-phase regimes (OPR).
For this reason, let us briefly consider the issue of stability of the in-phase mode. For this, we linearize the system~\eqref{eq:ModelGeneral} in the neighborhood of $\phi(t)$, representing $\varphi_n(t)$ in the form $\varphi_n\left(t\right)=\phi\left(t\right)+\delta\varphi_n\left(t\right)$. The linearized system of equations for perturbations $\delta \varphi_n(t)$ has the form:
\begin{equation}
	\begin{gathered}
	\delta\ddot{\varphi}_1+\lambda\delta\dot{\varphi}_1+\cos\phi\left(t\right)\delta\varphi_1=K\left(\delta\varphi_{2}-\delta\varphi_{1}\right),\\
	\delta\ddot{\varphi}_n+\lambda\delta\dot{\varphi}_n+\cos\phi\left(t\right)\delta\varphi_n=K\left(\delta\varphi_{n-1}-2\delta\varphi_n+\delta\varphi_{n+1}\right),\quad n=2,\ldots,N-1,\\
	\delta\ddot{\varphi}_N+\lambda\delta\dot{\varphi}_N+\cos\phi\left(t\right)\delta\varphi_N=K\left(\delta\varphi_{N-1}-\delta\varphi_{N}\right).
	\end{gathered}
	\label{eq:PerturbedChainLocal}
\end{equation}
We pass in the system ~\eqref{eq:PerturbedChainLocal} to the normal coordinates $\psi_1, \psi_2, \ldots, \psi_N$ and obtain the following system of independent equations~\cite{Bolotov-20}:
\begin{equation}
\ddot{\psi}_n+\lambda\dot{\psi}_n+\left[\cos\phi\left(t\right)-K\mu_n\right]\psi_n=0,
\label{eq:NormalMode}
\end{equation}
where $\mu_n=-2\left[1+\cos\left(n\pi/N\right)\right]$~\cite{Munyaev-20}. In our previous works~\cite{Bolotov-19, Bolotov-20, Smirnov-16} it was shown that the modes $\psi_1, \psi_2, \ldots, \psi_{N-1}$ can become unstable, which leads to the appearance of out-of-phase rotational modes, while the boundaries of the instability interval $(K_1^{\left(n \right)}, K_2^{\left(n \right)})$ are determined by the expressions
\begin{equation}
K_{1,2}^{\left(n\right)}=K_{1,2}^{*}/\left|\mu_n\right|,
\label{eq:K12n}
\end{equation}
where $K_{1,2}^{*}\left(\lambda,\gamma\right)$ are, respectively, the left and right boundaries of the region of instability of the in-phase solution of the equation $\ddot{\psi}+\lambda\dot{\psi}+\left[\cos\phi\left(t\right)+K^{*}\right]\psi=0$, which can be determined by the asymptotic expressions
\begin{equation}
K_{1,2}^{*}=\frac{1}{4}\left[\frac{\gamma^2}{\lambda^2}\mp2\sqrt{1-\gamma^2}+\frac{1}{2}\frac{\lambda^2}{\gamma^2}\right]+O\left(\frac{\lambda^4}{\gamma^4}\right).
\label{eq:K12c}
\end{equation}
Thus, in the system~\eqref{eq:ModelGeneral}, for certain values of the control parameters $\lambda$, $\gamma$, an $N-1$ region of instability of the in-phase mode $\phi(t)$ can exist. Moreover, these instability regions do not overlap at low dissipation ($\lambda \ll 1$).
\section{Intersection of in-phase regime  instability regions}~\label{sec:UnstabilityRanges}
\begin{figure}[htb]\center
	\includegraphics[height=0.3\columnwidth]{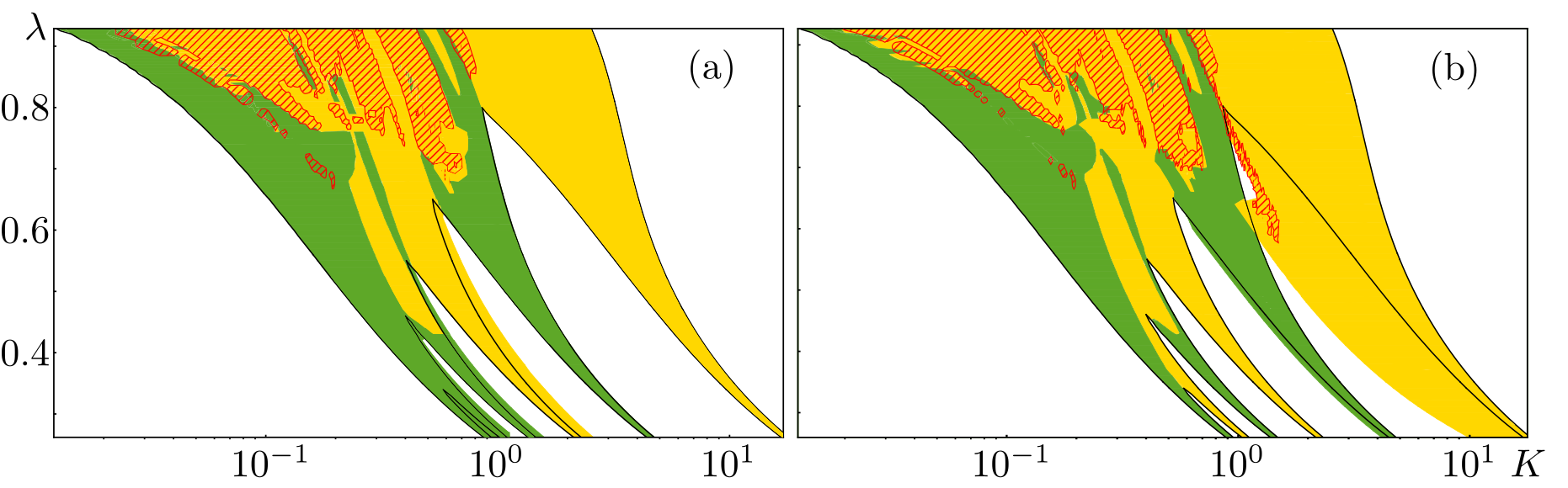} 
\caption{Maps of rotational modes are realized in the system~\eqref{eq:ModelGeneral} for $N=7$, $\gamma=0.97$ for different values of the $\lambda$ and $K$. Black lines -- boundaries of instability regions of in-phase rotational motion $\phi(t)$, obtained numerically. The color indicates the types of realized rotational modes: green -- $(2:2:2:1)$ regimes, yellow -- $(1:1:\ldots:1)$ regimes. Red shading indicates areas of chaos.	Maps are obtained by inheriting the initial conditions as a result of increasing (a) and decreasing (b) the parameter $K$.}
	\label{fig:7pend}
\end{figure}

In this section, we show that the mutual arrangement of the regions of instability of the in-phase regime significantly affects the complexity of the dynamic regimes being realized.
As a result of direct numerical simulation of the system~\eqref{eq:ModelGeneral} in the parameter area $\left(\lambda, \gamma \right)$, maps of rotational modes were built, shown in Fig.~\ref{fig:7pend},
analyzing which we see that there is a complication of structures with an increase of dissipation parameter. There is a convergence and overlap of the zones of various structures existence, as well as the emergence of multistability, which ultimately leads to the emergence of a regime of dynamic chaos.
\par
Let us analyze how the emergence, evolution and overlapping of the zones of instability of the IPR occurs. We show that for fixed values of the parameters $\lambda$ and $\gamma$ increased number of the chain elements $N$ leads to the following scenarios of overlap regions common-mode instability: 
\\
(i) one of the instability regions is separated from the rest when new elements are added; 
\\
(ii) there is such a critical number of elements $N^*$, when an exceeding instability region is separated from the rest; %
\\
(iii) for any number of elements $N$, all regions of instability have intersections with others.
\\
From the expression~\eqref{eq:K12n} and the monotonic increase of the expressions $\mu_1, \mu_2, \ldots, \mu_{N-1}$ it follows that the sequences of the left $K_1^{\left(1 \right)}, K_1^{\left(2 \right)}, \ldots, K_1^{\left(N-1 \right)}$ and right $K_2^{\left(1 \right)}, K_2^{\left( 2 \right)}, \ldots, K_2^{\left(N-1 \right)}$ boundaries of instability regions increase monotonically (the width of instability regions $K_2^{\left(n\right)}-K_1^{\left(n\right)}$ also increases monotonically). Thus, two near regions of instability can partially overlap ($K_1^{\left(n+1\right)}\le K_2^{\left(n\right)}$), or be separated by a stability window ($K_1^{\left(n+1\right)}>K_2^{\left(n\right)}$). Thus, the zone of instability cannot entirely enter into another zone of instability. Let us introduce the number $n^{*}\left(N \right)$, which determines the number of the first overlapping zones of instability with increasing bond strength parameter $K$. For example, if $n^{*}\left(8\right)=3$, then the first three zones of instability overlap, then the stability window follows (the location of the other four zones does not interest us, it is important that they are separated from each other). The expression $n^{*}\left(8\right)=7$ means that all zones of instability have overlapped and merged into one. It's obvious that $1\le n^{*}\left(N\right)\le N-1$. Formally, $n^{*}\left(N\right)$ is defined by the expression
\begin{equation}
n^{*}\left(N\right)=\min\left(\left\{n\in\left\{1,2,\ldots,N-2\right\}|K_1^{\left(n+1\right)}-K_2^{\left(n\right)}>0\right\}\cup\left\{N-1\right\}\right).
\end{equation}
To find an explicit expression for the number $n^{*}\left(N \right)$, consider the expression continualized by $n$ $K_1^{\left(n+1\right)}-K_2^{\left(n\right)}$, introduce the distance function between instability regions $L\left(x\right)=K_1^{\left(x+1\right)}-K_2^{\left(x\right)}$. Obviously, the function $L\left(x\right)$ is periodic: $L\left(x+2N\right)=L\left(x\right)$. In addition, there are gaps at the points $x=\left(2k+1\right)N$ and $x=\left(2k+1\right)N-1$. It can be verified that $\lim\limits_{x\to N}L\left(x\right)=-\infty$ and $\lim\limits_{x\to N-1}L\left(x\right)=+\infty$. Thus, the intervals $\left[-N,N-1\right]$ and $\left[N-1,N\right]$ contain at least one root of the equation $L\left(x\right)=0$. The direct solution of the equation $L\left(x\right)=0$ leads to the roots
\begin{equation}
x^{\pm}_k=\frac{2N}{\pi}\arctan\left(\frac{\cos\left(\frac{\pi}{2N}\right)\pm\sqrt{K_1^{*}/K_2^{*}}}{\sin\left(\frac{\pi}{2N}\right)}\right)+2Nk.
\end{equation}
The roots $x_0^{-}$ and $x_0^{+}$ are on the interval $\left[-N,N\right]$. We conclude that each interval $\left[-N,N-1\right]$ and $\left[N-1,N\right]$ contains one root, and the smaller root $x_0^{-}$ lies in the interval $\left[-N,N-1\right]$. Thus, for $x>x_0^{-}$ we have $L\left(x\right)>0$: the length of the gap between the regions of instability becomes positive, those zones no longer overlap. Then
\begin{equation}
n^{*}\left(N\right)=\max\left\{1,\quad\Bigg\lceil\frac{2N}{\pi}\arctan\left(\frac{\cos\left(\frac{\pi}{2N}\right)-\sqrt{K_1^{*}/K_2^{*}}}{\sin\left(\frac{\pi}{2N}\right)}\right)\Bigg\rceil\right\},
\label{eq:ncN}
\end{equation}
where $\lceil\ldots\rceil$ -- round up. Thus, as the parameter $K$ increases, only the first $n^{*}\left(N \right)$ instability regions can overlap. All other unstable regions (if they exist) are separated from each other by stability windows. It is also useful to note that due to the monotonic growth of the function under the round-up brackets in~\eqref{eq:ncN}, as the number $N$ grows, the sequence $n^{*} \left(N \right)$ is non-decreasing. In other words, an increase of the chain elements number $N$ can only lead to an increase in the number of intersecting regions of instability of IPR.
\par
Let us define the right boundary $K_2^{\left(n^{*}\right)}$ of the instability zone formed by the overlapping of the instability areas:
\begin{equation}
K_2^{\left(n^{*}\right)}=\frac{K_2^{*}}{4\cos^2\left(\frac{n^{*}\left(N\right)}{N}\frac{\pi}{2}\right)}.
\end{equation}
When $N\to\infty$, there is a limit $\lim_{N\to\infty}n^{*}\left(N\right)/N=1$. In fact, for sufficiently large $N$ (when $\cos\left(\frac{\pi}{2N}\right)>\sqrt{K_1^{*}/K_2^{*}}$)
\begin{equation}
n^{*}\left(N\right)=\Bigg\lceil\frac{2N}{\pi}\arctan\left(\frac{\cos\left(\frac{\pi}{2N}\right)-\sqrt{K_1^{*}/K_2^{*}}}{\sin\left(\frac{\pi}{2N}\right)}\right)\Bigg\rceil,
\label{eq:n*}
\end{equation}
the following inequalities are satisfied
\begin{equation}
\frac{2}{\pi}\arctan\left(\frac{\cos\left(\frac{\pi}{2N}\right)-\sqrt{K_1^{*}/K_2^{*}}}{\sin\left(\frac{\pi}{2N}\right)}\right)\le\frac{n^{*}\left(N\right)}{N}\le 1-\frac{1}{N}.
\end{equation}
Then, calculating the limits on both sides, by the two attendant theorem we arrive at the indicated limit.
As a result, we arrive at
\begin{equation}
\lim_{N\to\infty}K_2^{\left(n^{*}\right)}=+\infty.
\end{equation}
The left boundary of the instability zone formed by the overlap of individual instability regions is defined as $K_1^{\left(1\right)}$. When $N\to\infty$,
\begin{equation}
\lim_{N\to\infty}K_1^{\left(1\right)}=K_1^{*}/4.
\end{equation}
\par
For a more detailed study of the behavior of the instability regions we consider the number $\overline{n}^{*}\left(N\right)=N-1-n^{*}\left(N\right)$ ($0\le\overline{n}^{*}\left(N\right)\le N-2$), that determines the number of isolated (by stability windows) unstable regions. From the expression~\eqref{eq:ncN} we get that
\begin{equation}
\overline{n}^{*}\left(N\right)=\min\left\{N-1,\quad\Bigg\lfloor \frac{2N}{\pi}\left(\frac{\pi}{2}-\arctan\left(\frac{\cos\left(\frac{\pi}{2N}\right)-\sqrt{K_1^{*}/K_2^{*}}}{\sin\left(\frac{\pi}{2N}\right)}\right)\right)\Bigg\rfloor\right\}-1.
\label{eq:ntcN}
\end{equation}
Let's examine the behavior of the function under the round-down brackets. Direct computation can verify that its second derivative with respect to $N$ is always negative for $\forall N \ge 1$ (we are interested in the values $N\ge 2$). Thus, for $N \ge 1$, the first derivative decreases monotonically with increasing $N$. At the point $N=1$, the derivative is $\frac{K_1^{*}}{K_1^{*}+K_2^{*}}+\frac{2}{\pi}\arctan\sqrt{\frac{K_1^{*}}{K_2^{*}}}>0$; as $ N \to + \infty$, the value of the derivative tends to $+0$. Therefore it can be argued that the derivative all $N \ge 1$ is positive and of the function increases monotonically with an increase of the number $N \ge 2$. Then the sequence $\overline{n}^{*}\left(N\right)$, like the sequence $n^{*}\left(N\right)$, is non-decreasing.
\par
\begin{figure}[htb]\center
	\includegraphics[width=1.0\columnwidth]{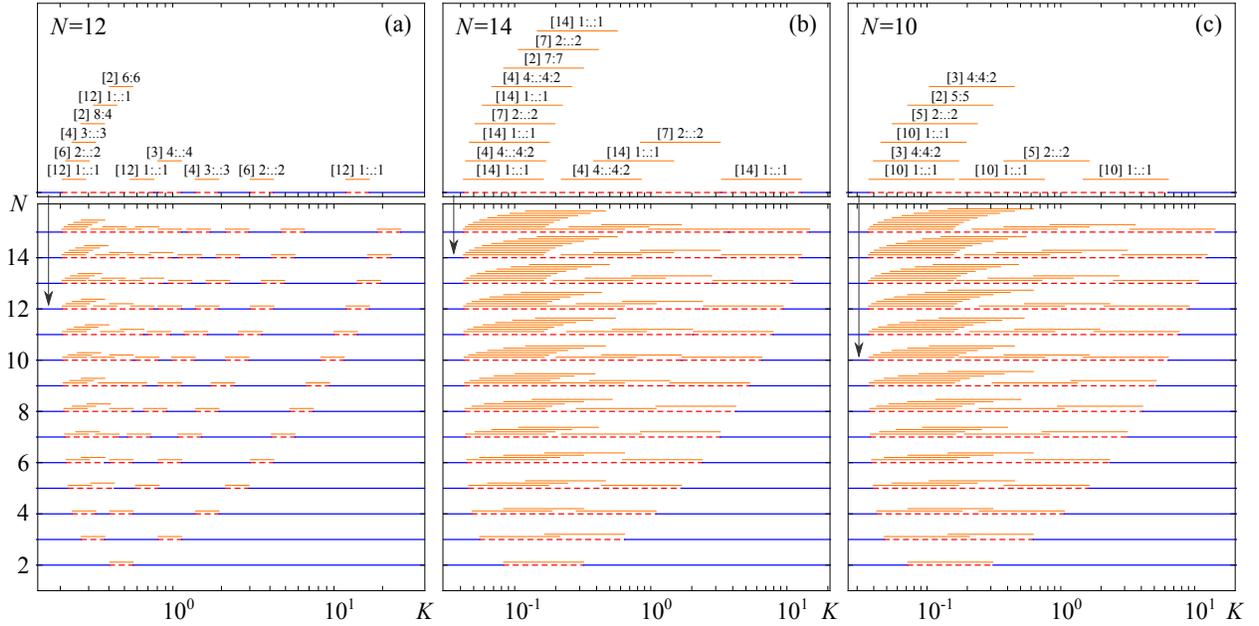}
	\caption{Lower panel. The regions of instability of the $\psi_n$ modes (orange solid lines), and the corresponding regions of instability of the IPR  (red dashed lines), found from the expression \eqref{eq:K12n} and the numerical calculation of the boundaries $K_{1,2}^{*}$, for $\gamma = 0.97$ and different $N$ and $\lambda$ depending on the bond strength $K$. Blue solid lines represent stable IPR. 
	Over panel. Instability regions of $\psi_n$ modes and the corresponding types of OPR for indicated by arrow number $N$. The number in square brackets indicates the number of in-phase clusters.
	(a) $\lambda=0.5$
	For any $N$, there is at least one region of in-phase instability, isolated from the others by stability windows.
	(b)$\lambda=0.8$. Since some $N$ (here $N=6$) appears common mode instability region isolated from other stability windows.
	(c)  $\lambda=0.82$. For any number of elements $N$, overlap of all instability regions of the $\psi_n$ modes is observed with the formation of instability area of the in-phase mode without stable windows.
	Over panel. Instability regions of $\psi_n$ modes and the corresponding types of out-of-phase rotation at $N=15$. The number in square brackets indicates the number of in-phase clusters.
	}
	\label{fig:InstabArea_0_5}
\end{figure}

{\it Scenario A.} Let us find a condition under which an isolated region of instability of IPR exists for any $N$. From the non-decreasing sequence $\overline{n}^{*}\left(N\right)$ it follows that a necessary and sufficient condition is the fulfillment of the equality $\overline{n}^{*}\left(3\right)=1$. From \eqref{eq:ntcN} we find $K_2^{*}/K_1^{*}<3$. The example of this scenario is demonstrated in Fig.~\ref{fig:InstabArea_0_5} (a).
\\
{\it Scenario B.} Similarly, one can find a condition under which all regions of instability overlap for any $N$. A necessary and sufficient condition is the equality
$\lim\limits_{N\to+\infty}\overline{n}^{*}\left(N\right)=0$. From \eqref{eq:ntcN} 
it follows $\lim\limits_{N\to+\infty}\overline{n}^{*}\left(N\right)=\bigg\lfloor\left(1-\sqrt{K_1^{*}/K_2^{*}}\right)^{-1}\bigg\rfloor-1$. 
Then we get $K_2^{*}/K_1^{*}>4$ (Fig.~\ref{fig:InstabArea_0_5}(b)). 

{\it Scenario C.}
When $3<K_2^{*}/K_1^{*}<4$ for small $N$ all instability regions overlap, but starting from some number $N$ isolated regions appear separated by stability windows (Fig.~\ref{fig:InstabArea_0_5}(c)).
\\
Another conclusion from the non-decreasing sequences $n^{*}\left(N\right)$ and $\overline{n}^{*}\left(N\right)$ is the fact that when adding a new element, i.e. as $N$ increases by one, either one new isolated region of instability appears and the number of overlapping regions does not change, or the number of overlapping regions increases by one, but the number of isolated regions does not change.
\clearpage
\section{The chaotic dynamics development}~\label{sec:chaos}
Here we analyze in detail the complex periodic and chaotic regimes occurrence mechanisms, and the various OPRs   existence regions overlapping role.
Below we present the results of computations. First, we introduce the synchronicity parameter
\begin{equation}
\Xi=\dfrac{1}{N\left(N-1\right)}\sum\limits_{n_1,n_2=1}^{N}\smash{\displaystyle\max_{0\leq t\leq T}}\left|\dot{\varphi}_{n_1}\left(t\right)-\dot{\varphi}_{n_2}\left(t\right)\right|,
\end{equation}
which characterizes the rotational motion phase synchronization degree. The value $\Xi=0$ shows that the rotational regime under consideration is in-phase, the values $\Xi>0$ indicate the realization of OPR. For a more detailed analysis, we present bifurcation diagrams, as well as graphs that show the local maxima of the oscillator frequencies and the largest Lyapunov exponent, which positive value indicates the dynamic chaos regime presence in the system.
\par
Let us investigate the rotational regimes dynamics in the chain depending on the parameter $K$ for different dissipation parameter $\lambda$ values and different elements number $N$. Our computational experiments show that with a parameter $K$ value increase the IPR  loses its stability softly, but the corresponding OPR stability loss with a further increase in $K$ occurs in a hard manner. Thus, hysteresis occurs when there is a range of $K$ values where in-phase mode coexists with out-of-phase mode. We will call such an interval of in-phase regime instability as ``right''. If the in-phase mode loses its stability hardly and the out-of-phase mode loses its stability softly with $K$ increasing then we will call such an instability range ``left''. The scenario for the appearance and existence of ``left'' and ``right'' regions for a different chain elements number $N$ is as follows:\\
$N=2$ -- one ``left'' $(1:1)$ region;\\
$N=3$ -- one ``left'' $(2:1)$, one ``right'' $(1:1:1)$;\\
$N=4$ -- two ``left'' $(1:1:1:1)$ and $(2:2)$, one ``right'' $(1:1:1:1)$;\\
$N=5$ -- two ``left'' $(2:2:1)$ and $(1:1:1:1:1)$, two ``right'' $(2:2:1)$ and $(1:1:1:1:1)$; etc.
\par
Thus, the transition from an even number of elements to an odd number is accompanied by the addition of one ``right'' area; similarly, from an odd number to an even number one ``left'' area is added.
\par
\begin{figure}[htb]\center
	\includegraphics[width=0.8\columnwidth]{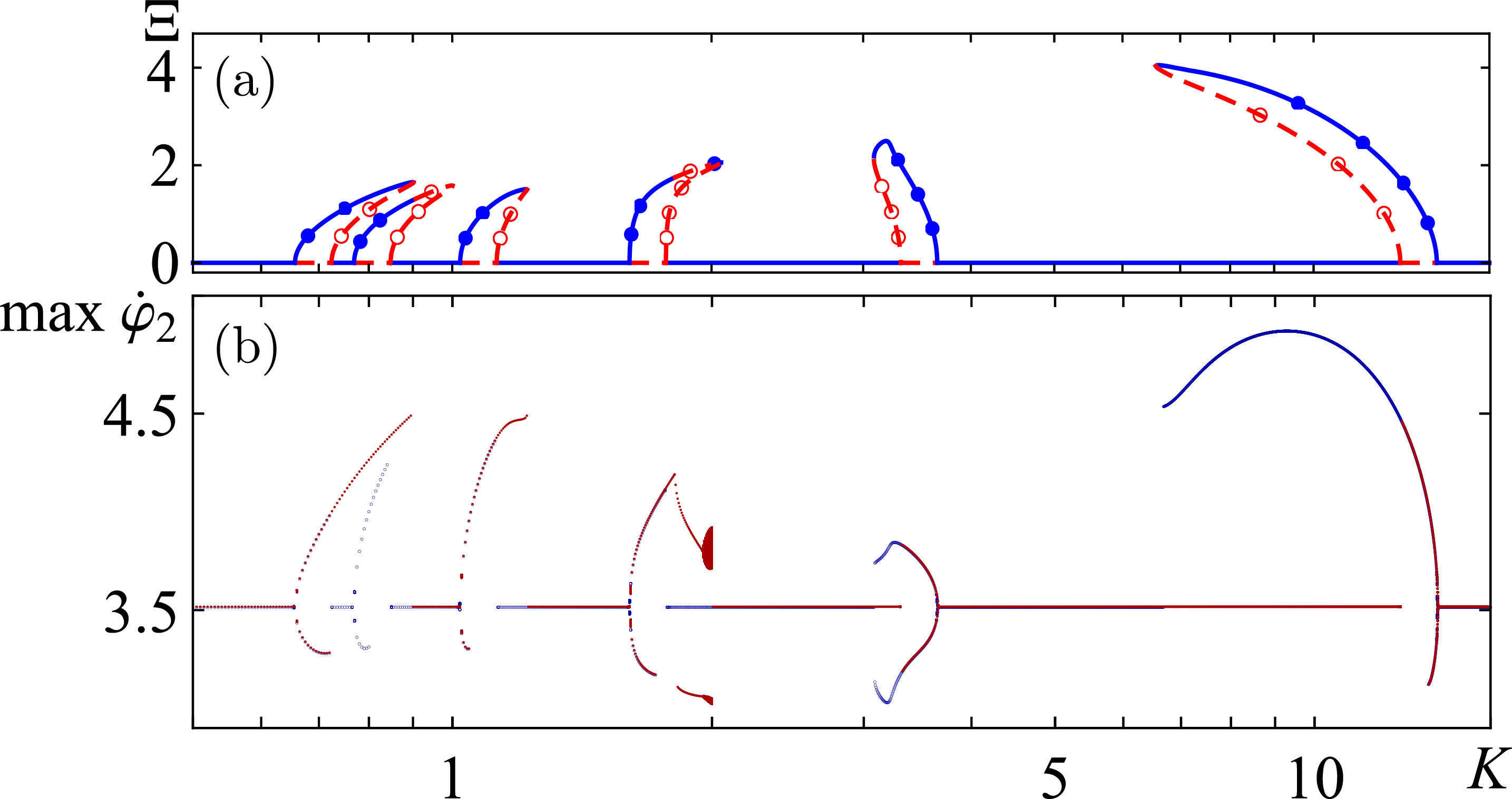}
\caption{Local bifurcation diagram of periodic rotational regimes (a). $\Xi$ -- synchronicity parameter. Round markers show $4\pi$-periodic rotational regimes. Filled markers correspond to stable rotational regimes, unfilled markers -- to unstable ones. The line without markers corresponds to the in-phase $2\pi$-periodic rotational regime, the solid line, to the stable one, and the dotted line, to the unstable one. Diagram is obtained in two ways: with increasing and decreasing parameter $K$. Local frequency maxima $\max\dot{\varphi}_m$ (b),
Parameters: $N=7$, $\gamma=0.97$, $\lambda=0.3$, $m=2$.}
\label{fig:bif07_03}
\end{figure}

Fixed number of elements  $N$. Consider the chain dynamics for $N=7$, $\gamma=0.97$, $\lambda=0.3$. In this case, with the parameter $K$ value increase the out-of-phase $(2:2:2:1)$ and $(1:1:\ldots:1)$ regimes associated with the in-phase rotation instability development and corresponding to the $\psi_n$ ($n=1,\ldots,6$) modes are sequentially realized in the system~\eqref{eq:ModelGeneral}. Note that modes with the same designation can differ from each other and represent different objects in the system~\eqref{eq:ModelGeneral} phase space. Consider the range of the parameter $K$ values when the rotational motion $(2:2:2:1)$ is realized for the first time. Here, as the parameter $K$ increases, the in-phase periodic rotation $\phi(t)$ undergoes a period-doubling bifurcation at $K\approx 0.65689$. In this case, stable in-phase $2\pi$-periodic motion gives rise to stable $4\pi$-periodic $(2:2:2:1)$ motion, but $\phi(t)$ loses its stability. The bifurcation diagram (Fig.~\ref{fig:bif07_03}) shows that there is also an unstable motion $(2:2:2:1)$, which at $K\approx 0.72420$ arises from an unstable in-phase motion as the period doubling bifurcation result, while $\phi(t)$ becomes stable again. Further, as the parameter $K$ increases, stable and unstable $(2:2:2:1)$ motions merge and disappear as a result of saddle-node bifurcation. With a further increase in the coupling parameter $K$ the regimes $(1:1:\ldots:1)$ and $(2:2:2:1)$ appear in a soft manner, and then regimes $(1:1:\ldots:1)$, $(2:2:2:1)$, $(1:1:\ldots:1)$ appear in a hard manner. Thus, when the parameter $K$ changes there are three ``right'' and three ``left'' in-phase regime $\phi(t)$ instability regions in the system~\eqref{eq:ModelGeneral} with small dissipation $\lambda$. Note also that for $\lambda=0.3$ $4\pi$-periodic rotational regimes $(2:2:2:1)$ and $(1:1:\ldots:1)$, referring to ``right'' instability intervals, can coexist on the interval of the coupling strength $0.76917\le K\le 0.90117$, although the corresponding instability regions do not intersect.

Let us analyze further the dissipation parameter $\lambda$ influence on the space-time dynamics complication in the system under consideration. As $\lambda$ increases, the arising out-of-phase rotations can also undergo the following period-doubling bifurcations, leading to the appearance of $8\pi$, $16\pi$, $32\pi$, etc. regimes. For example, for $\lambda=0.3$ the mode undergoes several period doubling bifurcations on the interval $1.81268<K<1.98930$ $(1:1:\ldots:1)$ (Fig.~\ref{fig:bif07_03}).

\begin{figure}[htb]\center
	\includegraphics[width=0.8\columnwidth]{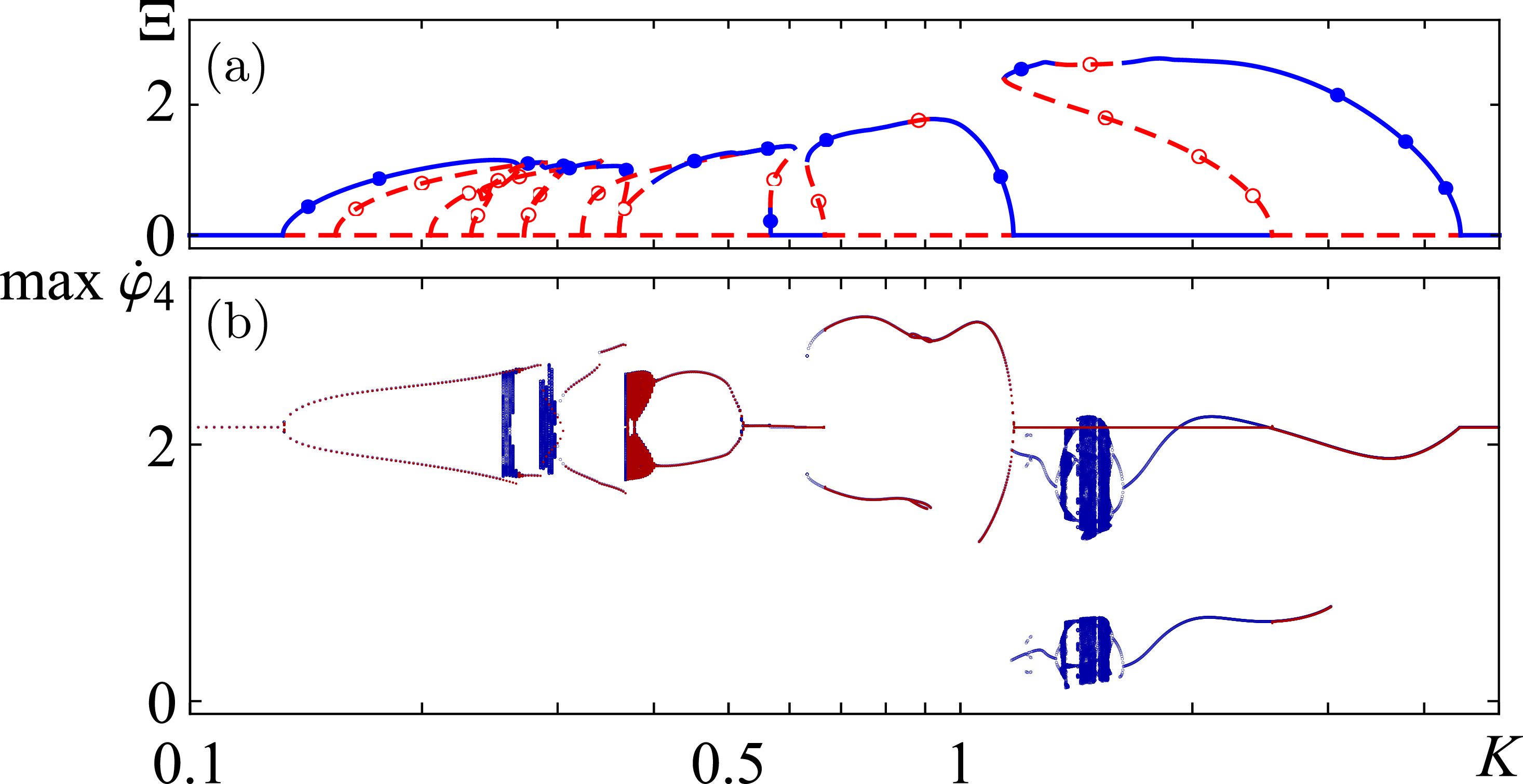}
	\caption{The same as in Fig.~\ref{fig:bif07_03}.  Parameters: $N=7$, $\gamma=0.97$, $\lambda=0.6$, $m=4$.}
	\label{fig:bif07_06}
\end{figure}

\begin{figure}[htb]\center
		\includegraphics[width=0.8\columnwidth]{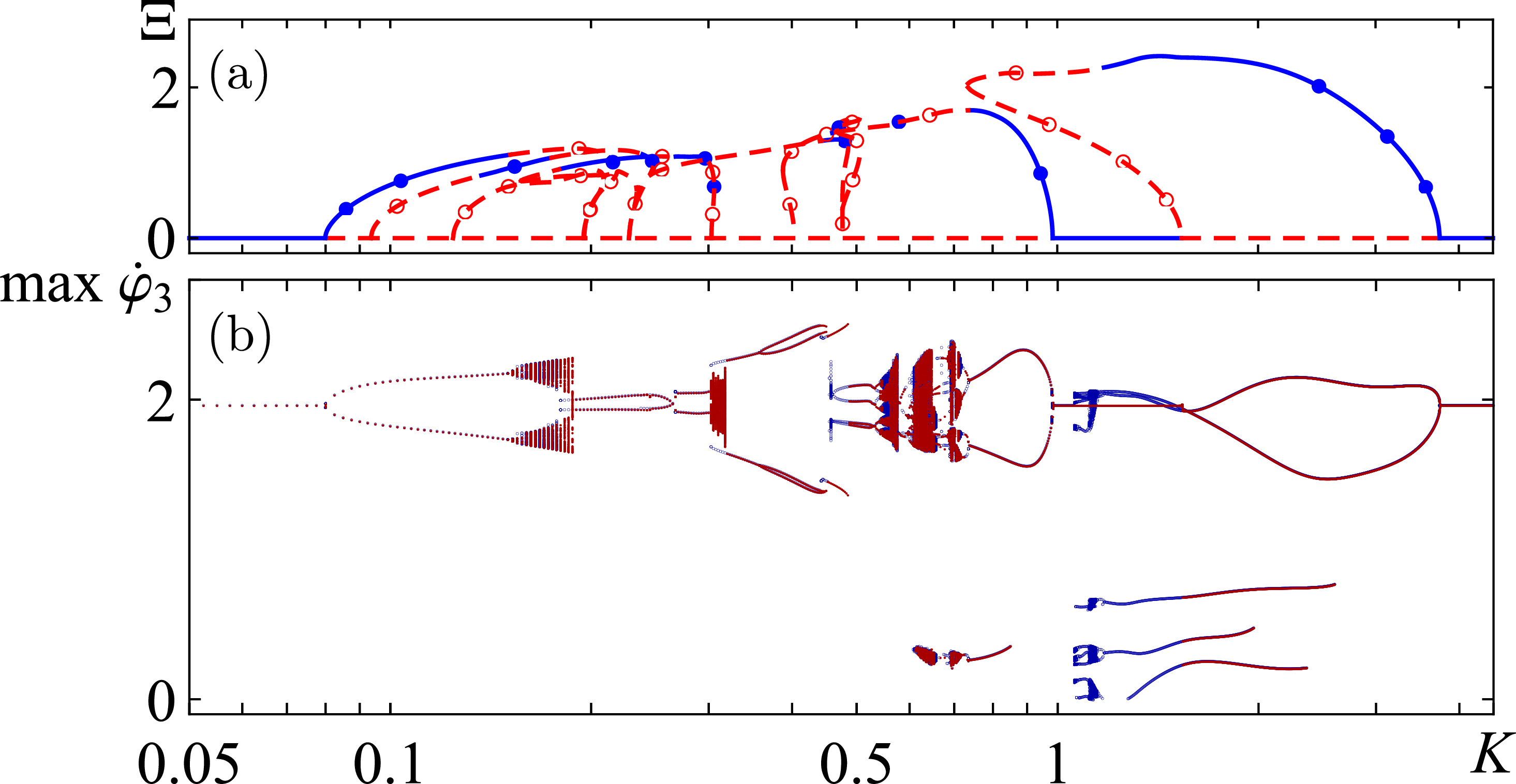}
	\caption{The same as in Fig.~\ref{fig:bif07_03}. Parameters: $N=7$, $\gamma=0.97$, $\lambda=0.7$, $m=3$.}
	\label{fig:bif07_07}
\end{figure}

\begin{figure}[htb]\center
		\includegraphics[width=0.8\columnwidth]{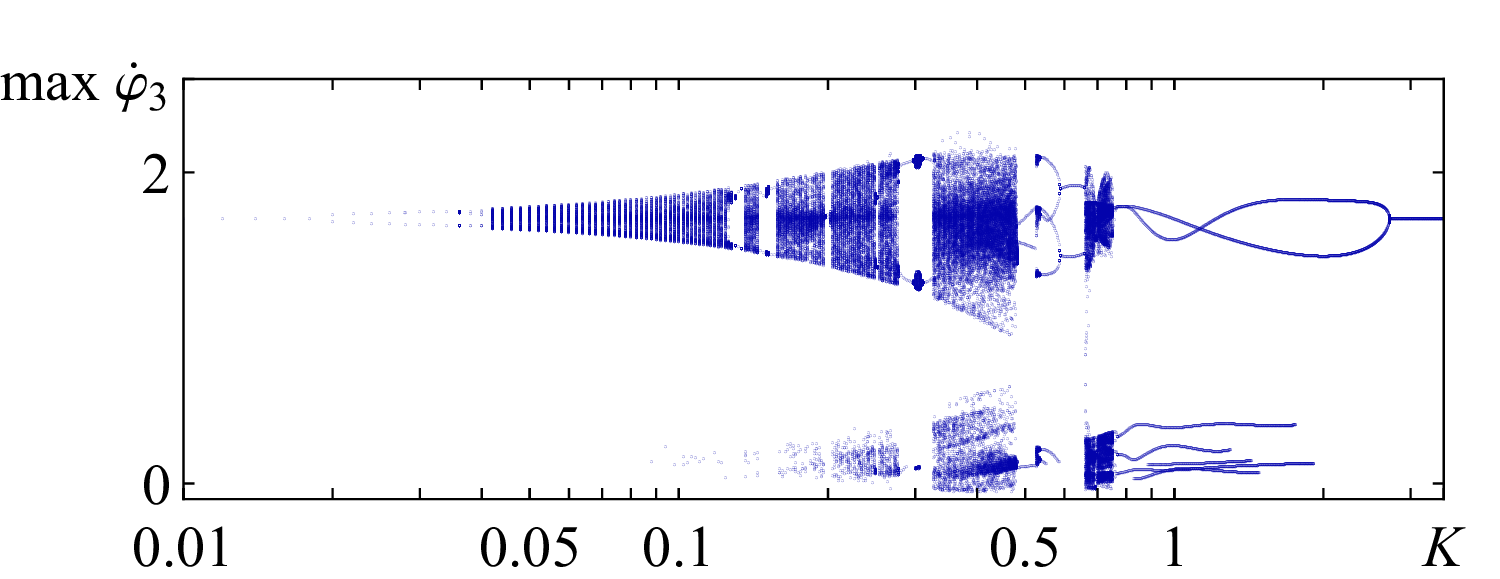}
	\caption{Local frequency maxima $\max\dot{\varphi}_m$. Parameters: $N=7$, $\gamma=0.97$, $\lambda=0.9$, $m=3$.}
	\label{fig:bif07_09}
\end{figure}

A further increase in the dissipation parameter $\lambda$ leads to the in-phase mode instability intervals intersections appearance. For $\lambda=0.6$ four regions corresponding to the smallest coupling parameter $K$ values intersect (Fig.~\ref{fig:bif07_06}), while the $4\pi$-periodic motions corresponding to the $(1:1:\ldots:1)$ regime are unstable and rotations with a larger number of turns by $2\pi$, which arise during subsequent $4\pi$-periodic trajectories bifurcations, are realized. In the ``left'' region corresponding to $(1:1:\ldots:1)$ regime with chaotic motions appears as a result of a period doubling bifurcations cascade, what is confirmed by the positive values of the largest Lyapunov exponent. In the case $\lambda=0.7$, when five instability regions intersect, chaotic regimes are observed already at several intervals. The transition from periodic to chaotic dynamics can occur through the Neimark-Sacker bifurcation leading to the emergence of an invariant torus, and its subsequent destruction through the torus destruction bifurcation which ultimately leads to the chaotic attractor birth. Here, the curves corresponding to $4\pi$-periodic rotational regimes have a rather nontrivial structure with different closures with each other (Fig.~\ref{fig:bif07_07}). With a further increase in the dissipation parameter $\lambda$ the regions of chaotic dynamics gradually increase and merge with one another (Fig.~\ref{fig:bif07_09}).

Fixed dissipation parameter. Let us consider the regimes evolution in the case of the fixed dissipation parameter value, for example, $\lambda=0.9$, with a change in the chain elements number $N$ and the coupling parameter $K$. First of all, we will be interested in the transitions from regular to chaotic regimes and the role played by the appearance and evolution of the IPR instability regions during these transitions. 

\begin{figure}[htb]\center
		\includegraphics[width=1.0\columnwidth]{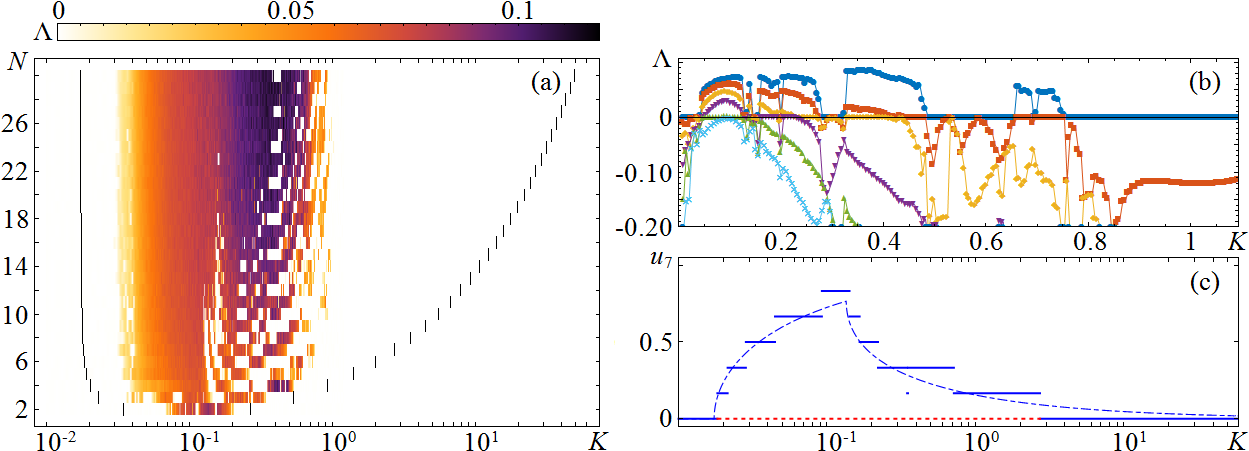}
	\caption{(a) Chaotic regimes map in the system~\eqref{eq:ModelGeneral} depending on the coupling strength $K$ and the elements number $N$ at $\gamma=0.97$ and  $\lambda=0.9$. Black lines denote the left $K_1^{(1)}$ and right $K_2^{(N-1)}$ boundaries of the IPR  $\phi(t)$ instability range. White color denotes regular rotational modes (largest Lyapunov exponent $\Lambda=0$). In colored  area the largest Lyapunov exponent is positive. (b) The spectrum (first six maximal)  Lyapunov exponents. There are intervals with four, three, two and one positive exponents. (c)   
	Dependence of the unstable modes $\psi_n$  for $N=7$ (blue lines) and for $N\to\infty$ (blue dashed curve). The red dashed line indicates the in-phase regime instability region for $N=7$.}
	\label{fig:chaos_09}
\end{figure}

Using the general expression \eqref{eq:K12n}, we can determine the fraction $u_N\left(K\right)$ of modes $\psi_n$ that lose their stability for some fixed value of the coupling parameter $K$:
\begin{equation}
u_N\left(K\right)=\frac{\theta\!\left(K\!-\!K_1^{*}/4\right)}{N\!-\!1}\Bigg\lfloor\!\frac{2N}{\pi}\arccos\!\left(\!\sqrt{\frac{K_1^{*}}{4K}}\right)\!\Bigg\rfloor-\frac{\theta\!\left(K\!-\!K_2^{*}/4\right)}{N\!-\!1}\Bigg\lfloor\!\frac{2N}{\pi}\arccos\!\left(\!\sqrt{\frac{K_2^{*}}{4K}}\right)\!\Bigg\rfloor,
\end{equation}
where $\theta$ is the Heaviside step function. Since $K^{*}_2/K^{*}_1>4$, then for any $N$ all instability regions of $\psi_n$ modes overlap, which leads to the formation of a single, global in-phase rotation instability region.  Find out what happens in the limit $N\to\infty$. For this case, we introduce the fraction $u_{\infty}\left(K\right)$ determined by the expression
\begin{equation}
u_\infty\left(K\right)=\frac{2}{\pi}\left[\theta\!\left(K\!-\!K_1^{*}/4\right)\arccos\!\left(\!\sqrt{\frac{K_1^{*}}{4K}}\right)-\theta\!\left(K\!-\!K_2^{*}/4\right)\arccos\!\left(\!\sqrt{\frac{K_2^{*}}{4K}}\right)\right].
\label{eq:u_inf}
\end{equation}
The   Fig.~\ref{fig:chaos_09} (c) demonstrates the dependence of the unstable modes fraction on the coupling parameter $K$ for fixed $N=7$ and in the limit $N\to\infty$.  The figure shows that the largest unstable modes number is observed at small and intermediate coupling parameter values ($0.01<K<1$). As a result of the IPR  stability loss new OPR appear, which in turn can also become unstable, which leads to the new and new OPRs emergence (see Figs.~\ref{fig:bif07_03}-\ref{fig:bif07_09}).

The general property of the obtained dependence \eqref{eq:u_inf} is its maximum at the point $K=K_2^{*}/4$, which is demonstrated in Fig.~\ref{fig:chaos_09}c, obtained for the considered case $\lambda=0.9$. Thus, the maximum $\psi_N$ mode instability regions overlapping density is observed at small and intermediate $K$ values located near the left boundary of the in-phase rotation instability interval, i.e. where the ``left'' instability regions appear.
\par
Our computational experiments show that chaotic regime precisely observed in the area of greatest accumulation and intersection of various instability regions. Fig.~\ref{fig:chaos_09} (a)  shows the area where chaotic dynamics is observed on the $(K,N)$ plane at $\lambda=0.9$. It can be seen that although the right boundary $K_2^{(N-1)}$ of the instability region moves rather quickly to the right as $N$ increases, chaotic regimes are realized only in a rather narrow range of coupling strength values $0.01<K<1$. At small $K$, due to weak interaction, chaotization of initially regular rotations does not occur. In the case of strong couplings, the pendulums interaction leads to the rotations regularization. In this case, the resulting modes are not necessarily in-phase. The in-phase rotational mode is established only with a large coupling strength (see the $K_2^{(N-1)}$ curve in Fig.~\ref{fig:chaos_09} (a)).
\par
Analyzing Fig.~\ref{fig:chaos_09}, the following conclusions can be drawn. For a fixed value of $K$ with a change in the chain length $N$, two main scenarios of the chaotic behavior emergence and existence are realized:\\
(i) for small $K \sim 0.1$, the chaos appearing at $N=2$ does not disappear with an increase in $N$;\\
(ii) for $0.12 \lesssim K \lesssim 1.0$: a) the addition of one (sometimes two or more) new elements can lead to both chaos and regularization of modes; b) when two (sometimes more) elements are added, the chaotic regime leaves it chaotic, and the regular regime remains regular. The parity of the elements number in the chain plays an important role here. In this case, the closer $K$ is to 1, the less dense of the overall chaos area becomes, i.e. the regular behavior islands number and size increase. For $K>1$, no chaotic behavior was found for any $N$.

A series of computational experiments was carried out in which the chain length $N$ varied with time. In one series, the number of elements was changed by adding (or excluding) one element. In the second series, several elements were added to the chain (excluded from the chain). In all cases, both of the above scenarios were observed.

For fixed $N$, the regular and chaotic regimes evolution with increasing $K$ is qualitatively the same for different chain lengths $N$. Namely, chaos appears practically at the same value of $K\approx 0.01$ and is realized continuously without regular behavior windows up to $K\approx 0.12$. Further, the chaotic regime can alternate with the regular one.

In studied chain of pendulums there is the oppotunity of existence of hyperchaos. We computed the spectrum of Lyapunov exponents in the chain of length $N=7$ and $\gamma=0.97$, $\lambda=0.9$ in dependence on coupling strength $K$  (Fig.~\ref{fig:chaos_09}(b)). There are several intervals of $K$ there four, three, two and one Lyapunov exponents are positive.   
The areas of existence of stable hyperchaotic regimes is strongly correlated with the areas of the largest number of unstable modes $\psi_n$.

\clearpage
\section{Conclusion}\label{sec:Conclusion}
The paper investigates the appearance and disappearance of a chaotic rotational regime in a chain of locally coupled identical pendulums. The discovered scenarios of the space-time chaos emergence in the ensemble under consideration are typical both for low-dimensional lumped dynamic systems and for multidimensional systems distributed over space. This is the birth of a chaotic attractor: a) through a period doubling bifurcations sequence of periodic motions; b) through the invariant tori destruction. The chaos appearance study was carried out with a change in: a) the dissipation parameter of an individual element; b) the coupling parameter and c) the interacting elements number. It is shown that an increase in dissipation in an ensemble with a fixed value of the coupling and the elements number can lead to the chaos appearance. This is due to the fact that with an increase in the dissipation parameter $\lambda$ (this leads to an approach to the Tricomi curve), the individual pendulum rotations become significantly inhomogeneous in time: intervals of fast and slow changes in the phase $\phi$ can be distinguished. A chaotic rotational regime can arise when even two such pendulums interact. Chaos in an ensemble can arise when the interacting elements number changes. Since in this case the control parameter is discrete, the chaos occurrence is rigid. With a fixed dissipation and coupling parameters values and with a change in the chain length $N$ two main scenarios of the chaotic behavior emergence and existence are realized: a) the occuring chaos at $N=2$ does not disappear with increasing $N$; b) the addition of one (sometimes two or more) new elements can lead to both chaotization and regularization of modes, and when two (sometimes more) elements are added, the chaotic regime leaves chaotic and the regular regime remains regular. In this case, the parity of the chain elements number plays an important role.

The coupling strength influence on the chaos occurrence is specific. For any chain lengths considered in this paper, the region of chaos existence is bounded and for $N>10$ does not depend on $N$. This is the range of small and intermediate coupling $0.01\lesssim K\lesssim 1.0$. The chaos existence in this range is due to the various out-of-phase rotational regimes existence regions overlap. There are no chaotic regimes were found in the chain with a strong coupling. It is important to note, that with increase of number of coupled elements the transition to hyperchaotic behavior is possible.  
\section*{Acknowledgments}
The numerical calculations in this work were supported by the Russian Science Foundation (grant No.~19-12-00367) and the analytical studies were supported by the Ministry of Science and Higher Education of Russian Federation (project No.~0729-2020-0036).

\end{document}